\documentclass[12pt,a4paper,english]{article}

\usepackage[latin9]{inputenc}
\usepackage{amstext}
\usepackage{graphicx}
\usepackage{esint}

\makeatletter

\newcommand{\lyxaddress}[1]{
	\par {\raggedright #1
	\vspace{1.4em}
	\noindent\par}
}
\newenvironment{lyxlist}[1]
	{\begin{list}{}
		{\settowidth{\labelwidth}{#1}
		 \setlength{\leftmargin}{\labelwidth}
		 \addtolength{\leftmargin}{\labelsep}
		 }}
	{\end{list}}

\@ifundefined{date}{}{\date{}}
\usepackage{babel}

\usepackage{babel}

\makeatother

\usepackage{babel}
\begin{document}

\title{Modeling micro-heterogeneity in mixtures: the role of many body correlations}

\author{Anthony Baptista and Aurélien Perera}
\maketitle

\lyxaddress{Laboratoire de Physique Théorique de la Matière Condensée (UMR CNRS
7600), Sorbonne Université, 4 Place Jussieu, F75252, Paris cedex 05,
France.}
\begin{abstract}
A two-component interaction model is introduced herein, which allows
to describe macroscopic miscibility with various modes of tunable
micro-segregation, ranging from phase separation to micro-segregation,
and in excellent agreement for structural quantities obtained from
simulations and the liquid state hypernetted-chain like integral equation
theory. The model is based on the conjecture that the many-body correlation
bridge function term in the closure relation can be divided into one
part representing the segregation effects, which are modeled herein,
and the usual part representing random many body fluctuations. Furthermore,
the model allows to fully neglect these second contributions, thus
increasing the agreement between the simulations and the theory. The
analysis of the retained part of the many body correlations gives
important clues about how to model the many body bridge functions
for more realistic systems exhibiting micro-segregation, such as aqueous
mixtures. 
\end{abstract}

\section{Introduction}

One of the central problem in statistical mechanics of liquids and
mixtures is to understand both qualitatively and quantitatively the
role of many-body effects on pair correlations\cite{01-0INTRO-mayer,02-0INTRO-Baxter}.
In its current state of development and application, liquid state
integral equation theory (IET) is mostly a theory formulated in terms
of pair interactions and pair correlations\cite{01-0INTRO-mayer,02-0INTRO-Baxter,03-0INTRO-hansmac}
, even though higher order correlations are incorporated implicitly
by the formalism itself and explicitly through the so-called bridge
term. Neglecting this term or using approximations involving only
pair correlations, leads to the usual strategies to generate approximate
IET\cite{01-0INTRO-mayer,02-0INTRO-Baxter,03-0INTRO-hansmac}. By
comparing pair correlations obtained from such approximate IET to
those obtained from computer simulations, one generally finds a rather
good agreement for various types of simple liquids\cite{04-0INTRO-caccamo},
which in turn suggests that the contribution from explicit many body
correlations remains quantitatively small, although being qualitatively
important. This agreement breaks down in the vicinity of phase transition,
such as the gas-liquid phase separation for one component liquids,
or demixing transitions for mixtures. This breakdown is intuitively
understood from the paramount importance of fluctuations through the
Renormalisation Group theory approach for liquids\cite{05-1RG-HRT,06-1RG-lomba},
indicating that many body correlations become essential in such cases,
although they globally remain quantitatively small\cite{06-1RG-lomba}.
Interestingly, approximate IET also break down when considering realistic
liquids which exhibit micro-structure\cite{07-1-faraday}. Such liquids,
which we call complex liquids, encompass associating liquids, such
as water and other hydrogen bonding liquids, and mixtures which show
micro-segregation, the latter for which it is often impossible to
solve IET\cite{07-1-faraday,07-2-dmso}.

From these observations it is tempting to postulate that the breakdown
of approximate IET is always caused by special types of spatial fluctuations,
such as critical ones or those related to spatial heterogeneity, which
reflects the importance of very specific types of many body correlations,
and which cannot be described by the usual approximations for IET.
To be more specific, approximate IET may be able to solve for micro-segregated
systems, but not with the required accuracy or quite simply fail beyond
a certain point, precisely because very specific forms of the many-body
contributions are required. It is these specific contributions, we
propose to consider as an effective interaction.  Remaining correlations
are considered as contributions from random fluctuations, whose importance
is similar to those in simple liquids, and therefore can be neglected
in approximate integral equation methods, since the principal many
body effects from non-random fluctuations are already captured through
the effective interactions.

The method of choice to investigate this approach is the hyper-netted
chain (HNC) approximation, which precisely consists in neglecting
contributions from high order correlation through the bridge term\cite{07-5morita,07-6virial,07-9fmtbridge}.
Interestingly, HNC is well known for having spurious tendency to fail
for the type of complex liquids mentioned above, which lesser approximate
IET do not have, although these do not necessarily provide solutions
in agreement with the expected ones\cite{07-rism,07-rism2}. This
failure is the starting point for the proposal in this work, which
consists in conjecturing that this shortcoming of HNC is the signature
that particular forms of the bridge function are required to describe
such systems.

Assuming these assumptions holds, it remain to find which particular
form of the bridge functions are required. Searching for clues, we
first note that HNC can handle fluctuations in several typical cases.
The first case concerns simple liquids and mixtures, such as Lennard-Jonesiums
or weakly polar liquids, for which there is no particular local order.
For such systems, the HNC approximation is fairly good\cite{04-0INTRO-caccamo},
and it is for this type of systems that more accurate alternative
empirical approximations have been developed, such as the Verlet bridge\cite{08-0B-verlet}
and many other methods\cite{04-0INTRO-caccamo,08-1B-rogerYoung,08-3B-hmsa,08-6B-labik}.
The second case concerns classes of liquids which present a strong
local order which dominates the typical disorder of the liquid state.
One such example concerns ionic melts, such a molten NaCl for example,
which are characterised by charge order\cite{09-1CO,09-3CO,09-5CO,09-6CO},
where positive and negative charges are disposed in quasi-alternate
fashion.  HNC is often found to describe the structure of such liquids
very accurately, even to extremely high couplings\cite{09-Ng}. With
these two considerations in mind, we note that all micro-segregating
systems are governed by strong Coulomb interactions, and it is precisely
these interactions that produce the segregation\cite{11-5MH}. Therefore,
if HNC is very good for some Coulomb systems and not for others, it
means that the bridge function of the latter systems must have a very
specific form and role, confirming the ideas developped above. Another
example concerns the so-called core-softened interactions\cite{10-1CS,10-2CS,10-3CS},
which aim to describe particular forms of local order, such as that
found in globular clustering\cite{10-1CS} or in water\cite{10-2CS}.
Among this category, the repulsive core-softened interaction is particularly
well described by the HNC approximation and in excellent agreement
with simulations, both for pattern formations\cite{10-5CSt2D} or
water-like models\cite{10-6CSwaterCS}. Since these models concerns
non-ionic systems, They give an indirect clue about the type of pseudo-interactions
which could mimic local order similar to that induced by Coulomb interactions,
without having to explicitly take them into account. This can be achieved
through the closure relation, as shown in the next section.

In this work, we wish to test these ideas through simple isotropic
pair interactions, but which contain the bridge part which concerns
the formation of micro-structure. If we formulate properly the effective
pair interactions, that is the concerned bridge part, we should able
to describe the whole expected scenario of micro-structure, as seen
from various experimental conditions. In addition, a proper formulation
should allow to neglect contributions from random fluctuations, and
in such case the HNC approximation should be highly reliable. The
several cases we wish to model concern the type of micro-segregation
found in aqueous mixtures. For example, aqueous mixtures of small
amphiphile molecules such as alcohols, are known to produce micro-segregation\cite{11-1MH,11-2MH,11-5MH},
which induces large pre-peaks in atom-atom structure factors\cite{11-5MH}.
IET are generally unable to provide solutions for such mixtures (as
in water-1propanol) \cite{12-DO}, or when they can, the description
of the structure is very poor (as in water-methanol)\cite{07-1-faraday}.
In addition to the pre-peak witnessing domain-domain correlations,
the $k=0$ behaviour of the structure factor is equally critical.
For example, a large value would indicate large domain fluctuations,
and not necessarily an underlying phase separation. These examples
further support the idea that micro-heterogeneity is a different form
of fluctuation than those which control phase transitions.

In the next section, we reformulate the conjecture in a more formal
way, and we explain the modeling and computational details. In section
III, we show the results of various modeling strategies, which allow
to extract key features of the effective interactions. In the last
section, we will discuss how these results can be extended to more
realistic case, before we conclude.

\section{Theoretical and computational details}

\subsection{From many body correlations to effective interactions}

To be more specific about the statements of the Introduction, let
us consider the most general form\cite{03-0INTRO-hansmac,04-0INTRO-caccamo}
of the pair distribution function $g_{ab}(r)$ for a mixture of atomic
sites, interacting through spherically symmetric pair interaction
$v_{ab}(r)$:

\begin{equation}
g_{ab}(r)=\exp\left(-\beta v_{ab}(r)+h_{ab}(r)-c_{ab}(r)+b_{ab}(r)\right)\label{clos}
\end{equation}
where the index $a$ and $b$ designate atomic species in the mixture,
$h_{ab}(r)=g_{ab}(r)-1$, $c_{ab}(r)$ is the pair direct correlation
function, and $\beta=1/k_{B}T$ is the Boltzmann factor (with $k_{B}$
the Boltzmann constant and $T$ the temperature). The last function
$b_{ab}(r)$ is the so-called bridge function\cite{07-5morita,07-9fmtbridge}
which contains all the many body higher order correlations, namely
through the introduction of n-body direct correlation functions $c_{a_{1}a_{2}...a_{n}}^{(n)}(\vec{r}_{1},\vec{r}_{2},...,\vec{r}_{n})$,
where the index $a_{i}$ is a species index, $(n)$ designating the
rank or order of correlations and $\vec{r}_{i}$ designating the position
of particle $i$ of species $a_{i}$: 
\begin{equation}
b_{ab}(r)=\sum_{m\geq3}\frac{1}{m!}b_{ab}^{(m)}(r)\label{b}
\end{equation}
with 
\begin{equation}
b_{ab}^{(m)}(r)=\sum_{k\geq3}^{m}\frac{1}{k!}\sum_{s_{1}...s_{k}}\left(\prod_{l=1}^{k}\rho_{s_{l}}\right)T_{ab\,s_{1}...s_{k}}^{(k)}(r)\label{b1}
\end{equation}
and

\begin{equation}
T_{ab\,s_{1}...s_{k}}^{(k)}(r)=\int d\vec{r}_{13}\int d\vec{r}_{14}...\int d\vec{r}_{1k}\left(\prod_{l=1}^{k}h_{as_{k}}(r_{1k})\right)c_{bs_{1}...s_{n}}^{(k+1)}(\vec{r}_{2},\vec{r}_{3},...,\vec{r}_{k})\label{b2}
\end{equation}
where we have used the isotropy and translational invariance in macroscopically
homogeneous liquid mixtures, with $r=|\vec{r}_{12}|$ , $r_{ij}=|\vec{r}_{ij}|=|\vec{r}_{j}-\vec{r}_{i}|$.
This simplification has been omitted in the argument of the n-body
direct correlation function, in order to avoid explicitly counting
all combinations of $r_{ij}$, but it is obviously implied.

There is a striking complexity level difference between Eq.(\ref{clos})
and those defining $b_{ab}$(r). Tentative to compute low order contributions
to $b_{ab}(r)$, either through Mayer bond\cite{07-6virial} or directly
from Eqs.(\ref{b}-\ref{b2}) by approximating the function $c^{(3)}$
\cite{13-c3}, have indicated that the density expansion is diverging.
This has been directly confirmed by the exact computation of series
expansion for the case of 1-dimensional hard sphere fluid. It is therefore
hopeless to evaluate $b_{ab}(r)$ term by term, and one should model
this term directly by some general guiding rules. This is what approaches
such as the Verlet bridge\cite{08-0B-verlet} and all its descendants\cite{04-0INTRO-caccamo,08-1B-rogerYoung,08-3B-hmsa,08-6B-labik}
have tried to do. In the present work, we propose a different route,
based on the existence of the two categories of local order described
in the Introduction.

The conjecture formulated in the Introduction can be written as:

\begin{equation}
b_{ab}(r)=b_{ab}^{(LF)}(r)+b_{ab}^{(RF)}(r)\label{bnew}
\end{equation}
where $b_{ab}^{(LF)}(r)$ corresponds to many-body contributions to
the local fluctuations (LF), while $b_{ab}^{(RF)}(r)$ corresponds
to random fluctuations (RF). For simple liquids one has

\begin{equation}
b_{ab}(r)\approx b_{ab}^{(RF)}(r)\hspace*{1em}\hspace*{1em}\text{simple liquids}\label{bsimple}
\end{equation}
and it is this term that are usually approximated in methods designated
to improve IET.

The lowest approximation level in the diagrammatic expansion in $b_{ab}(r)$
consist in setting $b_{ab}(r)=0,$ which is known as the hypernetted-chain
approximation (HNC) 
\begin{equation}
g_{ab}(r)=\exp\left(-\beta v_{ab}(r)+h_{ab}(r)-c_{ab}(r)\right)\label{hnc}
\end{equation}
Other levels of approximation involved further manipulation of the
HNC closure (such as the PY or MSA approximations\cite{03-0INTRO-hansmac}),
or adhoc functional expression for $b_{ab}(r)$ which are based on
various empirical basis (such as the Verlet approximation and the
large number of variants based on it). Since the HNC closure represents
the lowest approximation level based on the rigorous developments,
we will exclusively consider this approximation below, deliberating
neglecting various other approximations which are sometimes considered
as more accurate from various types of empirical considerations.

In cases where complex local order is present, that is when particles
tend to form local patterns (such as chaining in the case of dipolar
molecules\cite{14dipolechain}, for example), we postulate that the
influence of the $b_{ab}^{(LF)}(r)$ term cannot be neglected, and,
therefore, the HNC approximation is inappropriate. This is true in
particular for aqueous mixtures which exhibit micro-segregation. In
order to account for such local order, one must guess a functional
form for the $b_{ab}^{(LF)}(r)$ term, while the $b_{ab}^{(RF)}(r)$
term can be neglected in a first approximation. We postulate that
$b_{ab}^{(LF)}(r)$ must have generic functional forms, and can be
modeled properly if such forms are known. With this idea in mind,
the new form of the HNC approximation can be rewritten as 
\begin{equation}
g_{ab}(r)=\exp\left(-\beta\tilde{v}_{ab}(r)+h_{ab}(r)-c_{ab}(r)\right)\label{nhnc}
\end{equation}
where 
\begin{equation}
\beta\tilde{v}_{ab}(r)=\beta v_{ab}(r)-b_{ab}^{(LF)}(r)\label{veff}
\end{equation}
represents the effective interaction which accounts for the local
order. One justification for this rewriting is the success of some
core-soft models, for which we consider that the second outer core
is in fact a representation of $-b_{ab}^{(LF)}(r)$. Another class
of models which justify the rewriting concerns the so-called SALR
1-component models\cite{30-0SARL,30-1SARLR,30-2SALR,30-4SALR} (SALR
stands for short range attraction, long range repulsion), where the
long range repulsion part accounts for the $-b_{ab}^{(LF)}(r)$ and
which helps stabilize the local clustering in these models.

\subsection{From effective interactions to the $b^{(LF)}(r)$ bridge function}

In this part, we propose to determine the $b_{ab}^{(LF)}(r)$ term
by using pseudo-potentials instead of the original interactions which
produce the micro-segregation and the relevant complex disorder. In
order to do that, we first observe that complex disorder is invariably
produced by strong orientational interactions, and more particularly
hydrogen bonding interactions for the cases we are concerned with.
In the classical force field approach, hydrogen bonding is described
through Coulomb pairing interactions and partial charges $z_{a}$
on selected atomic sites $a$, such as oxygen, hydrogen and nitrogen
atoms, for example. Typical classical force field used in computer
simulation of molecular liquids are of the form: 
\begin{equation}
v_{ab}(r)=v_{ab}^{(LJ)}(r)+v_{ab}^{(C)}(r)\label{v(r)}
\end{equation}
where $v^{(LJ)}(r)=4\epsilon_{ab}\left[\left(\sigma_{ab}/r\right)^{12}-\left(\sigma_{ab}/r\right)^{6}\right]$
is the standard (12-6) Lennard-Jones pair interaction, and $v_{ab}^{(C)}(r)=z_{a}z_{b}e^{2}/r$
is the Coulomb pair interaction.

From the statistical theory of liquids, it is well known that the
pair direct correlation functions is related to the pair interaction
$v_{ab}(r)$ at large separations, through the exact relation\cite{03-0INTRO-hansmac}:

\begin{equation}
\lim_{r\rightarrow\infty}c_{ab}(r)=-\beta v_{ab}(r)\label{limc}
\end{equation}
This relation is used to handle numerically the long range part of
the Coulomb interaction $v_{ab}^{(C)}(r)$ in a convenient way\cite{09-Ng},
by separating out the short range part $c_{ab}^{(SR)}(r)$ of $c_{ab}(r)$
from a long range part which is handled analytically through an error
function\cite{09-Ng}:

\begin{equation}
c_{ab}(r)=c_{ab}^{(SR)}(r)-A_{ab}\frac{\mbox{erf}(\alpha_{ab}r)}{r}\label{Ccoul}
\end{equation}
where $A_{ab}=-z_{a}z_{b}\beta e^{2}$ (where $z_{i}$ is the valence
of atom $i$ and $e$ is the elementary charge), and $\alpha_{ab}$
is chosen such that it is smaller than the particle diameter $\sigma_{ab}$.
This latter point is very important since it means that, for distances
larger than $\alpha_{ab}$ , $r>\alpha_{ab}$, the error function
part of $c_{ab}(r)$ in Eq.(\ref{Ccoul}) exactly cancels the corresponding
Coulomb interaction $v_{ab}^{(C)}(r)$ in the closure relation Eq.(\ref{clos}).
This implies that the closure equation can be exactly rewritten as:
\begin{equation}
g_{ab}(r)=\exp\left(-\beta v_{ab}^{(LJ)}(r)+h_{ab}(r)-c_{ab}^{(SR)}(r)+b_{ab}(r)\right)\label{closSR}
\end{equation}
where only the short range interactions remain. While this might be
a useful trick\cite{09-Ng}, the closure in Eq.(\ref{closSR}) is
rigorously equivalent to Eq.(\ref{clos}) for distances outside particles
overlap such that $g_{ab}(r)\neq0$. Eq.(\ref{closSR}) is very interesting
in our case because it allows to understand the role played by the
bridge term $b_{ab}(r)$. Indeed, if we consider a system interacting
solely through LJ type interactions, then Eq.(\ref{closSR}) indicates
that, whatever the features attesting the existence of micro-segregation
in the pair correlation functions $g_{ab}(r)$, $h_{ab}(r)$ and $c_{ab}^{(SR)}(r)$,
these features can arise only through the term $b_{ab}(r)$, which
now plays the role of an supplementary pseudo potential. If we remove
this term, then we are back to simple disordered LJ liquids where
no micro-segregation or clustering can appear. Eq.(\ref{closSR})
is the formal proof that one could consider pair distribution functions
$g_{ab}(r)$ taken from a micro-segregated system and find $b_{ab}(r)$
which will produce the same pair correlations for the effective interaction
$\tilde{v}_{ab}(r)=v_{ab}^{(LJ)}(r)-b_{ab}(r)/k_{B}T$, now defined
in terms of the non-Coulomb part of the pair interaction. Conversely,
if the appropriate $b_{ab}(r)$ are injected as pseudo-interactions
in an ordinary LJ mixture, it will turn it into a micro-segregating
mixture, with the same pair distribution functions $g_{ab}(r)$. In
this perspective, Eq.(\ref{closSR}) now becomes a strict HNC closure,
which can be solved in conjunction with the Ornstein-Zernike (OZ)
equation\cite{03-0INTRO-hansmac} for the pseudo-system. Obviously,
the corresponding direct correlation $c_{ab}(r)$ will differ from
the solution of the original system, since it does not obey the same
OZ equation which involves the full Coulomb interactions. 

We can use the above equivalence to our advantage to extract the $b_{ab}^{(LF)}(r)$
term which we seek. We can use toy interactions which produce clustering
and micro-segregation in a model mixture, which produces pair distribution
functions $g_{ab}(r)$ similar to those found in realistic mixtures,
and we can solve HNC for such models and see how well we reproduce
the desired features in $g_{ab}(r)$. The pseudo potential is supposed
to contain a standard repulsion-attraction part $v_{ab}^{(0)}(r)$,
which handles the core and the dispersive parts of the interactions,
and a supplementary part $v_{ab}^{(S)}(r)$ which accounts for the
complexity (clustering, micro-segregation): 

\begin{equation}
\tilde{v}_{ab}(r)=v_{ab}^{\mbox{0}}(r)+v_{ab}^{(S)}(r)\label{veff-new}
\end{equation}
The results above suggest that $b_{ab}(r)=-\beta v_{ab}^{(S)}(r)$
is the bridge term we seek. In practice however, the HNC solution
for the pseudo-system may not exactly match that of the simulated
pseudo-system, because of the bridge term related to the pseudo-system
itself. The key assumption of this paper is that the most appropriate
analytical form of pseudo-potential $v_{ab}^{(S)}(r)$ would match
the local fluctuation part of the bridge function $b_{ab}^{(LF)}(r)$
introduced in Eq.(\ref{bnew}), even though it may not match the full
bridge function. The differences between the $g_{ab}(r)$ obtained
from HNC and the real model system would then be attributable to either
incorrect modeling of the pseudo-potential $v_{ab}^{(S)}(r)$, or
to the physical importance of the random fluctuation part of the bridge
function $b_{ab}^{(RF)}(r)$ in Eqs.(\ref{bnew},\ref{bsimple}).
This can be tested only through trial and error, as shown in the Results
section.

We can now reformulate the expression for the bridge function we seek
as:

\begin{equation}
b_{ab}^{(LF)}(r)=-\beta v_{ab}^{(S)}(r)\label{bridge}
\end{equation}
which, together with Eq.(\ref{veff-new}) is the principal result
of this paper.

Summarising our strategy, instead of considering the full model, which
creates a specific form of local order, we mimic such system by projecting
the many-body $-b_{ab}^{(LF)}(r)$ term into the effective interaction.
This way, using the HNC approximation in Eq.(\ref{nhnc}), we freeze
the local ordering by imposing it directly into the pair interactions,
and expect that we can neglect the contributions of the random fluctuations
by setting $b_{ab}^{(RF)}(r)=0$ . It is important to realise that,
when adopting this attitude, the usual control parameters such as
the temperature and the density do not have the same meaning. Indeed,
the effective interaction term $-b_{ab}^{(LF)}(r)$ is supposed to
depend on such parameters, and change when these are varied. Instead,
one should \emph{adapt} newer forms of the full effective interaction
$\tilde{v}_{ab}(r)$ to the expected local order which depends on
different temperatures and densities.

\subsection{Structure factors}

We monitor the desired forms of order by looking at their manifestations
in the pair correlation functions $g_{ab}(r)$ as well as in the corresponding
structure factors $S_{ab}(k)$ 
\begin{equation}
S_{ab}(k)=\delta_{ab}+\rho\sqrt{x_{a}x_{b}}\int d\vec{r}\left[g_{ab}(r)-1\right]\label{Sab}
\end{equation}
We emphasize that, at this stage of the exploration, it is not essential
to compute thermo-physical properties, nor to expect that these would
agree well with the expected values (as obtained from computed from
simulations, for example). Indeed, these would still depend on the
magnitude of the neglected $b_{ab}^{(RF)}(r)$ terms. For these reasons,
we will focus solely on comparing $g_{ab}(r)$ and $S_{ab}(k)$ with
the corresponding values computed through the computer simulations.
Furthermore, in order to appreciate the relation between local order
and $b_{ab}^{(LF)}(r)$, we will try as much as possible to minimize
the $b_{ab}^{(RF)}(r)$. In other words, we wish to model $\tilde{v}_{ab}(r)$
in such a way that the HNC approximation is quasi-exact for the corresponding
model system.  In this way, we expect to find universal forms of
the bridge functions with account for the micro-segregation in more
realistic models. Indeed, it is important to note that the present
formulation considers atomic sites, but does not specify if these
belong to molecules or not. For example, in the site-site formulation
of molecular liquids\cite{15ssoz1}, the closure relation Eq.(\ref{clos})
does not contain any reference to the molecular nature. It is only
at the level of the Ornstein-Zernike (OZ) equation that the molecular
shape is introduced, for example through the W-matrix in the site-site
OZ equation\cite{15ssoz1,16ssoz2}. From this point of view, the present
formulation is perfectly general, and should be able to describe molecular
liquids as well, and in particular associating hydrogen bonding liquids. 

 In addition to the structure factors in Eq.(\ref{Sab}), which represent
species-species density correlations $S_{ab}(k)=<\rho_{a;\mathbf{k}}\rho_{b;\mathbf{-k}}>$,
where $\rho_{c;\mathbf{k}}$ is the Fourier transform of the instantaneous
microscopic density of species $c$, we monitor the so-called Bathia-Thornton
(BT) structure factors\cite{31-BT} which represent the correlations
between total microscopic density $\rho_{N;\mathbf{k}}=\rho_{1;\mathbf{k}}+\rho_{2;\mathbf{k}}$and
``concentration'' density $\rho_{C;\mathbf{k}}=x_{2}\rho_{1;\mathbf{k}}-x_{1}\rho_{2;\mathbf{k}}$.
The resulting BT structure factors are related to the previous structure
factors, the most interesting relations being:

\begin{equation}
S_{NN}(k)=\frac{1}{2}\left[S_{11}(k)+S_{22}(k)\right]+S_{12}(k)\label{SNN}
\end{equation}

\begin{equation}
S_{CC}(k)=x_{2}^{2}S_{11}(k)+x_{1}^{2}S_{22}(k)-2x_{1}x_{2}S_{12}(k)\label{SCC}
\end{equation}
These structure factors are helpful in the way they allow the interpret
k-dependent fluctuations in terms of global density variables instead
of species related variables, hence reflecting the global heterogeneity
of the system. $S_{NN}(k)$ is the equivalent of an effective 1-component
structure factor for the mixture, while $S_{CC}(k)$ reflects the
relative heterogeneity between the 2 components. This way, the BT
representation allows to separate the homogeneous and heterogeneous
components from the $S_{ab}(k)$ structure factors. 

\subsection{Models of the effective interactions}

As stated in the Introduction, micro-heterogeneity in realistic systems
is produced principally through Coulomb interactions, which tend to
create a strong local order, which necessitates explicit contributions
from high order correlations. In this work, we do not wish to address
directly the problem by explicitly taking into account Coulomb interactions.
Yet, we would like to capture the corresponding local heterogeneity.
To be more precise, we would like to capture the specific heterogeneity
which cannot be addressed by HNC in the realistic case. The central
idea is to consider that the realistic system itself introduces the
specific many-body correlations, in addition to the realistic interactions
that creates the micro-heterogeneity. This is somewhat similar to
going from the atomic representation of a molecule to the molecular
representation as an aggregate of atoms. The specific bridge term
$b^{(LF)}(r)$ are considered here to play the role of an ``intra-aggregate''
interaction. This way, we intend to reduce the initial problem to
that of considering only this reduced interaction. 

In order to model such pseudo-interactions, we rely on previous investigations,
such as the SALR interactions mentioned previously, which are known
to produce micro-segregation effects. In addition, we search for other
similar forms, but with the idea in mind that HNC should be sufficient
to describe the system, in other words $b^{(RF)}(r)\approx0$. This
way, we can compare the results of HNC with those of computer simulations
of systems with substituted interactions, in place of the realistic
ones. The task of testing the forms of the $b_{ab}^{(LF)}(r)$ for
different types of realistic interactions showing micro-segregation,
is relegated to subsequent investigations. 

The SALR one-component model, mentioned above, consists of the usual
short range repulsion and a long range attraction, both of which ensuring
a condensed liquid phase, followed by a weak long range repulsion,
which controls the formation of local aggregation in restricted parts
of the phase diagram. Such types of models have been intensely investigated
in the past\cite{30-0SARL,30-1SARLR,30-2SALR,30-4SALR}. The standard
SALR model contains 2 exponential in order to model the short range
attraction and the long range repulsion, and has the same form as
Eq(\ref{veff-new}):

\begin{equation}
\tilde{v}(r)=v_{0}(r)-\epsilon_{1}\sigma\frac{\exp\left[-(r-\sigma)/\kappa_{1}\right]}{r}+\epsilon_{2}\sigma\frac{\exp\left[-(r-\sigma)/\kappa_{2}\right]}{r}\label{SARL}
\end{equation}
where $v_{0}(r)$ is a bare interaction terms, which takes into account
the particle core and any dispersive interaction. This could be a
Lennard-Jones (LJ) interaction, or a soft sphere $1/r^{12}$ interaction.
In the remaining two other contributions, all the $\epsilon_{\alpha}$
parameters are positive and have the dimension of an energy, and $\sigma$
is the diameter of the particle. The first Yukawa interaction helps
aggregate particles, while the second Yukawa helps segregate the aggregates
formed by the first. Unfortunately, the one-component model neglects
the crucial role of the solvent in the auto-organisation process which
enforces the aggregation of the solute. For example, this is one of
the essential feature behind the hydrophobic effect\cite{10-3CS}
.

In the present work, we propose to extend this model to a 2 component
system, which consists in of both mono-atomic solvent, labeled 1,
and a solute, labelled 2, interacting through the usual short range
repulsion. We will consider only models where both the solvent and
the solute have the same size with a common diameter $\sigma_{11}=\sigma_{22}=\sigma$.
The cross diameter $\sigma_{12}$was initially left free, as to consider
non-additivity, but the final retained choice was also $\sigma_{12}=\sigma.$
Tests for $\sigma_{22}>\sigma_{11}$ did not alter the main conclusion
of this work as far as the working hypothesis behind Eq.(\ref{bnew})
is concerned, and will be reported elsewhere. The key feature is in
the cross-interaction modeling. The interactions are as follows

\begin{equation}
\tilde{v}_{11}(r)=\tilde{v}_{22}(r)=v_{0}(r)\label{SRLA1}
\end{equation}
\begin{equation}
\tilde{v}_{12}(r)=v_{12}^{(0)}(r)+v_{12}^{(\mbox{rep})}(r)+v_{12}^{(\mbox{att})}(r)\label{SRLA12}
\end{equation}
The cross-species interaction $\tilde{v}_{12}(r)$ contains, in addition
to the $v_{12}^{(0)}(r)$ term, two additional contribution, the first
$v_{12}^{(\mbox{rep})}(r)$ which repel particles of different species,
hence leading to macroscopic demixion, and the second $v_{12}^{(\mbox{att})}(r)$
which controls this demixing tendency, leading to micro-segregation.
This is very different from the SALR model, since it is exactly the
opposite tendency. We propose to call this new model SRLA, by analogy
with SALR. An alternate model to SRLA would have been to implement
the SALR mechanism for the solvent and the solute separately, leaving
the cross interaction neutral. However, this would require more adjustable
parameters, with no clear justification as how one would choose to
control the relative balance between repulsion and attraction for
each of the components. The SRLA model is the minimal model for the
purpose defined herein.

The desired bridge function $b_{12}^{(LF)}(r)$ is then given by Eq.(\ref{bridge}):

\begin{equation}
b_{12}^{(LF)}(r)=-\beta v_{12}^{(\mbox{rep})}(r)-\beta v_{12}^{(\mbox{att})}(r)\label{b12}
\end{equation}
while the like-bridge terms are set to zero: $b_{11}^{(LF)}=b_{22}^{(LF)}=0$,
since in the present formulation, the micro-segregation is supported
only by the cross interactions.

\subsection{Integral equation theory}

This theory consists in solving the Ornstein-Zernike and a closure
equation, for which we use here the HNC closure of Eq.(\ref{nhnc}),
in order to solve for the pair correlation functions $g_{ab}(r)$
and the associated direct correlation functions $c_{ab}(r)$ are used
to study these binary mixtures. For a mixture with $n$ number of
components, the OZ equation can be written and a $n\times n$ functional
matrix equation:

\begin{equation}
\mathbf{SM}=\mathbf{I}\label{OZ}
\end{equation}
with $S_{ab}=S_{ab}(k)$ as defined by Eq.(\ref{Sab}) and $M_{ab}$
defined as

\begin{equation}
M_{ab}(k)=\delta_{ab}-\rho\sqrt{x_{a}x_{b}}\tilde{c}_{ab}(k)\label{Mk}
\end{equation}
where $\tilde{c}_{ab}(k)$ is the Fourier transforms direct correlation
functions $c_{ab}(r)$.

In our case we consider binary mixtures with $n=2$. hence we solve
for the functions $g_{11}(r)$, $g_{12}(r)$ and $g_{22}(r)$, where
$1$ designates the first component (the solvent) and $2$ the second
component (the solute). In the usual practice, these functions are
discretized over an array of $2048$ points, with a distance step
of $\delta r=0.02\sigma$. This allows to use fast Fourier transformation
techniques to obtain the structure factors. We use a standard iterative
technique to obtain the solutions of the 2 equations.

\subsection{Computer simulations}

We have used an in-house Monte Carlo code to perform Canonical ensemble
simulations, with constant NVT. All simulations were made for an equimolar
mixture of N=4000 particles. For each system, 50000 equilibration
moves were initially done starting from a random mixture, and followed
by 50000 sampling runs for statistical properties. This last step
corresponds, in the usual formulation, to 200 millions Monte Carlo
statistics per state points. The pair correlation functions obtained
from such statistics are smooth enough to be directly compared with
the corresponding quantities obtained from the IET techniques described
above. We emphasize that the $g_{ab}(r)$ obtained in these simulations
require a shift of their asymptote to the expected value 1, since
it is now well known\cite{35-1LP,35-2LP} that finite size effects
irremediably affect these asymptotes by a shift factor $\epsilon_{ab}/N$
, where $N$ is the number of particle in the box and $\epsilon_{ab}$
is related to the concentration fluctuations within the box. We obtain
the shifting factor empirically by enforcing the running integral
$G_{ab}(r)=4\pi\int_{0}^{r}u^{2}du\left[g_{ab}^{(s)}(u)-1\right]$
to reach a flat asymptote, where $g_{ab}^{(s)}(r)=g_{ab}(r)/(1-\epsilon_{ab}/N)$
is the function corrected for the shift. Even though the functions
obtained in this work tend to oscillate at long range due to domain
correlations, the quantities $G_{ab}(r)$ usually tend to curve downward
(for like correlations) or upward (for unlike correlation). This is
corrected empirically through the factor $\epsilon_{ab}$ adjusted
to straighten these asymptotes. For the type of models studied herein,
in several cases, the shift was found to be negligibly small, and
in such cases the procedure was ignored.

\section{Results}

The most important feature we would like to reproduce is the micro-segregation
of the solvent and the solute, and the consequences of such micro-segregation
on the structural properties, which are \emph{i)} out-of-phase long
range domain oscillations between like correlations $g_{11}(r)$,$\;g_{22}(r)$
and cross correlations $g_{12}(r)$, ii ) subsequent domain pre-peaks
in the structure factors $S_{ab}(k)$, which are positive for $S_{11}(k)$
and $S_{22}(k)$ and negative for $S_{12}(k)$. These latter signs
are direct mathematical consequence of the fact that cross interaction
leads to depletion of species of opposite type, hence small short
range correlations for $g_{12}(r)$ around the contact region. We
wish to emphasize that correlation functions and structure factors
of micro-structured systems have these features which are absent from
the same quantities in simple liquids and mixtures. These features
are essentially due to many body correlations, and  approximate 
IET are not able to reproduce them, precisely because they lack contributions
from such correlations.

Since the hypothesis behind Eq.(\ref{bnew}) is to separate ``frozen''
many body correlations from random ones, it is important to appreciate
the relative balance between these 2 terms. The guiding factor behind
our investigation is the known fact, stated in the Introduction, that
HNC is less accurate for core-softened models with an attractive part
than with purely repulsive ones. The rationale behind this finding
is that the second core tends to order particles around the central
one by depleting their numbers, and the attractive component in the
interaction tends to counter this depletion effect, leading to increase
the probability of finding more particles around the central one,
hence increasing the number of possibilities. Translated in terms
of Eq.(\ref{bnew}), the influence of attractive interactions is to
increase the importance of $b_{ab}^{(RF)}(r)$, when we would prefer
to reduce it, in order to appreciate the importance of $b_{ab}^{(LF)}(r)$.

In the next first two subsections we study previous categories of
models, such as the SALR model or the non-additive model, both of
which have been used to model micro-segregation. We demonstrate that
none of these models allows to satisfactorily satisfy the hypothesis
behind this work; namely to allows to fully neglect the random fluctuations
contributions from many body correlations.

\subsection{Unsuccessful model: the Y-SRLA}

A first natural idea was to follow the SALR pathway and use Yukawa
interactions for $v_{12}^{(\mbox{rep})}(r)$ and $v_{12}^{(\mbox{att})}(r$)
in Eq.(\ref{SRLA12}) The resulting model we named Y-SRLA (with the
Y for Yukawa), leading to

\begin{equation}
v_{0}(r)=v_{12}^{(0)}(r)=4\epsilon_{0}\left(\frac{\sigma}{r}\right)^{12}\label{Y1}
\end{equation}

\begin{equation}
v_{12}^{(\mbox{rep})}(r)=\epsilon_{1}\sigma\frac{\exp\left[-(r-\sigma)/\kappa_{1}\right]}{r}\label{Y2}
\end{equation}

\begin{equation}
v_{12}^{(\mbox{att})}(r)=-\epsilon_{2}\sigma\frac{\exp\left[-(r-\sigma)/\kappa_{2}\right]}{r}\label{Y3}
\end{equation}
It turned out that this model could not produce satisfactory domain
segregation, such as to produce wide oscillatory domain correlations
in the long range. In addition the agreement between theory and simulation
was quite weak, and specially more so when we tried to increase the
domain correlations. The reason for this problem can be tracked to
an important feature of this Yukawa interaction, which lead us to
a change in strategy. The problem is the existence of the range parameters
$\kappa_{i}$, which in fact introduce an additional length scale
into the problem, in addition to the slow exponential decay. It turns
out that what seems required for domain oscillation is a not a range+decay,
but a \emph{localisation} of the interaction. We found this out by
substituting a Gaussian to the second exponential. We believe that
this is an important information concerning the nature of the frozen
many body correlations. They cannot be properly modeled by screened
Coulomb interactions which do not localize the particles very well.
In particular, it would seem that the second attractive interaction
requires to be localized such as a Gaussian.

\subsection{Unsuccessful model: the non-additive SRLA}

With the failure of the previous model, we turned our attention to
non-additive interactions (hence the name NA-SRLA). However, since
non-additivity is more toy model than one suited to describe realistic
systems, we have tested models where is is the cross interaction dispersive
interaction range which controls the long range re-mixing. In this
context, the non-additivity is brought by the short range repulsion
$\tilde{v}_{12}^{(\mbox{rep})}(r)$. The interaction considered in
this section are of the form:

\begin{equation}
v_{0}(r)=4\epsilon_{0}\left[\left(\frac{\sigma}{r}\right)^{12}-\lambda_{0}\left(\frac{\sigma}{r}\right)^{6}\right]\label{NA1}
\end{equation}
and the generic form for the cross interaction becomes:

\begin{equation}
\tilde{v}_{12}(r)=4\epsilon_{12}\left[\left(\frac{\sigma_{12}}{r}\right)^{12}-\lambda\left(\frac{\sigma_{12}}{r}\right)^{6}\right]+\epsilon_{2}\exp\left[-(r-\alpha_{1})^{2}/\kappa_{1}^{2}\right]\label{NA2}
\end{equation}
With this form, the extension of the repulsive part covers both first
terms in Eq.(\ref{SRLA12})

Fig.1(a-b) shows one such typical trial for the following parameters,$\epsilon_{0}=1$,
$\lambda_{0}=1$, $\epsilon_{12}=2.8$, $\lambda=1$, $\epsilon_{1}=40$,
$\alpha_{1}=0.75$, $\kappa_{1}=0.2582$, and for state parameters
temperature $T=1$ and packing fraction $\eta=(\pi/6)\rho=0.4$. The
pair interactions shown in the inset of Fig.1a shows how the repulsive
part of the cross interaction installs a pseudo non-additivity, while
at the same time allowing for a long range attraction through the
depth controlled by the large value of $\epsilon_{12}$.

In this figure, we can see that the domain oscillations do not develop
very well, although the corresponding positive and negative pre-peaks
at $k\sigma_{1}\approx2$ are quite apparent in Fig.1b. In addition,
the agreement with simulation remains qualitative, much like the one
found for the Y-SRLA model mentioned in the previous section.  The
BT structure factors shown in the lower inset of Fig.1b indicate that
the $S_{NN}(k)$ looks very much as a 1-component structure factor
of ordinary LJ liquid, while the $S_{CC}(k)$ shows a prominent pre-peak
corresponding to those of $S_{11}(k)$ and $S_{22}(k)$, hence witness
the clustering induced heterogeneity in the system, as expected from
the input pair interaction. This inset provides a direct illustrated
of the separation, mentioned in Section 2.1, of the homogeneous and
heterogeneous components of the $S_{ab}(k)$ structure factors, into
$S_{NN}(k)$ and $S_{CC}(k),$respectively. It also shows that while
the former is well reproduced by HNC, it is quite inaccurate in the
latter, as shown by the larger pre-peak HNC predicts.

In a second example in Fig.2(a-b), we considered purely repulsive
first term, in order to get rid of the fluctuations associated to
the short range attraction in the like interaction part, with the
expectation to get a better agreement between simulations and IET.
Fig.2(a-b) illustrate the results with parameters $\epsilon_{0}=1$,
$\lambda_{0}=0$, $\epsilon_{12}=1$, $\lambda=1$, $\epsilon_{1}=4$,
$\alpha_{1}=0.85$, $\kappa_{1}=0.4663$, and for state parameters
temperature $T=1$ and packing fraction $\eta=(\pi/6)\rho=0.4$

One important draw back of this model is that it was impossible to
get into the regime where the $k=0$ values where large, with smaller
pre-peaks, that were in good agreement with simulations. This is crucial
to describe near-demixing situations, where the cross LJ attraction
is too small compared to the short range repulsion which guides the
demixing. It would seem that, when we neared such conditions, the
differences between the (exact) simulation results and the theory
started to differ. This situation reminds what happens in simple liquids
near demixing transitions, where the contribution of fluctuations
do not allow the theory to work well. We traced back this situation
to the fact of keeping a LJ interaction for the first part, a form
which is remisniscent of the simple liquids, for which the theory
is not so good.  The upper inset of Fig.2b, as well as the BT structure
factors in the lower inset show that HNC overestimates the heterogeneity
in the $S_{CC}(k)$ pre-peak as compared with simulations. 

From the perspective of the conjecture formulated herein, the form
of the pseudo-potential proposed in Eq.(\ref{NA2}) seems to leave
a rather large part of the many body correlation description into
the $b^{(R)}(r)$ part mentioned in Eq.(\ref{bnew}), as witnessed
by the inappropriate reproduction of the pre-peak feature. Hence,
it makes this form not suitable for proper modeling of the heterogeneity
bridge function.

\subsection{Successful model: the G-SRLA}

In the final form, we settled for using Gaussian for both terms $v_{12}^{(\mbox{rep})}(r)$
and $v_{12}^{(\mbox{att})}(r)$, which turned out to be the correct
choice. It confirms the hints provided in sub-section 3.1, where the
localisation seems to be the keyword. The exact forms of the interactions
used for this study are:

\begin{equation}
v_{11}(r)=v_{22}(r)=4\epsilon_{0}\left(\frac{\sigma_{12}}{r}\right)^{12}\label{G1}
\end{equation}

\begin{equation}
\tilde{v}_{12}(r)=4\epsilon_{12}\left(\frac{\sigma}{r}\right)^{12}+\epsilon_{1}\exp\left[-(r-\alpha_{1})^{2}/\kappa_{1}^{2}\right]-\epsilon_{2}\exp\left[-(r-\alpha_{2})^{2}/\kappa_{2}^{2}\right]\label{G2}
\end{equation}
with generic parameters $\epsilon_{0}=\epsilon_{12}=1$, and for state
parameter temperature $T=3$ and packing fraction $\eta=(\pi/6)\rho=0.4$.
The parameters of the first Gaussian have been fixed at $\epsilon_{1}=0.5$,
$\alpha_{1}=1.5$, $\kappa_{1}=0.41$, and for the second Gaussian,
we fixed the parameters $\alpha_{2}=3$, $\kappa_{2}=0.41$ and varies
the depth $\epsilon_{2}$ of the attraction to describe various cases.
It is this depth which controls the segregation. In the following
figures, we illustrate how this form allows to describe the 3 typical
situations, ranging from near demixing to full domain segregation,
particularly visible in the shape of the structure factors in the
small-k region.

Fig.3(a-b) show the near demixing, with large $k=0$ values for the
structure factors. Demixing is enhanced when $\epsilon_{2}=0$, since
there are only repulsive cross interactions. We start with a small
$\epsilon_{2}$ parameter, namely $\epsilon_{2}=0.015$, in order
to see how much it helps re-entrant mixing. It is seen that this is
yet not similar to the usual demixing in simple liquids, since the
cross term is negative. It is seen that, although the agreement between
the calculated and simulated correlations are in close agreement in
the short to medium range part, there is a small difference in the
long range part, which leads to structure factors that differ quite
strongly near $k=0$. This is probably due to the hidden influence
of the random fluctuation part $b^{(RF)}(r)$, which we totally neglect
in the HNC theory.  This disagreement reveals that we cannot expect
the HNC theory to describe properly systems close to true phase transitions,
for the lack of knowing how critical fluctuations influence correlations.
Nevertheless, the agreement is much better than in the 2 previous
cases. The BT structure factors in Fig.3b show that the small-k raise
is not due to critical fluctuation, in which case it would appear
in $S_{NN}(k),$but to the heterogeneity, since it appears in $S_{CC}(k).$
This is an interesting demonstration that large heterogeneity can
be mistaken for critical concentration fluctuation, and that the BT
transformation can help figure out such cases. On the other hand,
it is clear that, when $\epsilon_{2}\rightarrow0$ in Eq.(\ref{G2})
then true critical demixing will occur because of the purely repulsive
cross interactions. Hence, Fig.3(a-b) appears as a show case for many
realistic aqueous mixtures, such as water-acetonitrile for example,
which show both considerable heterogeneity and demixing-like behaviour\cite{35-xN,35-yM}.

Fig.4(a-b) shows an intermediate case, for the parameter $\epsilon_{2}=0.030$,
which corresponds to increasing the depth of the attractive interaction
with respect to the previous case in Fig.3. This is a very interesting
case from physical point of view, since it concerns a system which
hesitates between demixing and micro-segregation. This is usually
called a Lifshitz point\cite{36-1Lifshitz}, and our model is capable
of describing this physical situation in excellent agreement with
simulations. The Lifshitz point appears in many circumstances in the
context of micro-emulsions, where there is a triple coexistence between
two homogeneous and a layered phase\cite{30-4SALR,36-1Lifshitz}.
But, in the present case, it describes the turning point between segregated
phase and a micro-segregated phase. Such phase was found in a realistic
mixture of water and diols in a previous study.

Fig.5(a-b) illustrates the case with full micro-segregation and domain
ordering. It corresponds to the parameter $\epsilon_{2}=0.12$.  We
note the quasi perfect agreement between HNC and simulations both
in the correlation and structure factors, particularly for the BT
ones.

Finally, Fig.6(a-b) described a case where we vary the Gaussian width
parameter in order to increase the region of micro-segregation. It
corresponds to parameters $\epsilon_{2}=0.025$ but we also modify
$\kappa_{2}=4.47$. This particular example shows very clearly how
domain oscillations appear in conjunction with a growth of the pre-peaks
in the structure factors witnessing domain-domain correlations. These
features look very similar to those we have reported in several types
of aqueous mixtures\cite{11-5MH,12-DO}. Similar results are equally
obtained when $\epsilon_{2}$ is kept fixed while varying $\kappa_{2}$.
This shows that the model is very flexible, in terms both height and
width of the second Gaussian, and more importantly allows to control
the extinction of the second bridge term $b_{ab}^{RF}(r)$, thus making
HNC an exact relation in terms of the effective interaction.

From the results above, we propose that the cross bridge term for
real clustering and micro-segregated system may be efficiently modeled
as a sum of two Gaussians:

\begin{equation}
b_{12}^{(LF)}(r)=-\beta\epsilon_{1}\exp\left[-(r-\alpha_{1})^{2}/\kappa_{1}^{2}\right]+\beta\epsilon_{2}\exp\left[-(r-\alpha_{2})^{2}/\kappa_{2}^{2}\right]\label{b2G}
\end{equation}

The application of the above expression for realistic systems remains
to be tested for various cases. We expect that the insertion of the
above expression in the pair correlation function between the principal
hydrogen bonding atoms of different species will to take into account
the main feature of the micro-segregation, and allow HNC to be finally
solve for such systems.

\section{Discussion }

In order to formulate the existence of the separation in Eq.(\ref{bnew})
we have implicitly accepted the existence of domain correlations in
complex liquids, in particular aqueous mixtures  , and more importantly
that such domains emerge a new form of interaction, which can be captured
through the bridge terms $b^{(LF)}(r)$, the lack of which will not
permit to find numerical solution with the HNC IET . Insights into
such correlations are obtained from our own previous investigations
into these systems. Long range domain oscillations were observed from
computer simulations of aqueous-1propanol\cite{12-DO} and also ethanol-benzene
mixtures\cite{35-2LP}. These oscillations simply reflect the local
segregation into nano-domains of each species.

In view of the fact that previous models such as SALR models have
been designed to mimic aggregates in pseudo-one component systems,
our approach provides a link between the domain segregation observed
in complex mixtures and the models such as SALR models. In Section
2.2 we have demonstrated that such effective models which capture
the complexity of the disorder (clustering, self-segregation, and
others), are in fact formulating explicit expressions for the many
body correlation bridge terms, in particular those related to local
fluctuations $b_{ab}^{(LF)}(r)$. In section 3, we have used this
property of effective models to provide explicit expressions for the
cross species bridge term $b_{12}^{(LF)}(r)$.

The study provided in Section 3.3 clearly indicates that the $k=0$
raise in the structure factor cannot be solely attributed to random
fluctuations such as those appearing near spinodal lines or critical
points. This is a very important point, and a hint to this problem
has been previously provided in the study of such system in the two-dimensional
cases\cite{40-1CO2D}.

The previous point does not exclude the fact that the $k=0$ raise
observed for simple liquids in the vicinity of mechanical instability
remains of point delicate to investigate through the IET methodology.
This means that systems which tend to organise locally and which are
submitted to large random fluctuations, will not be well described
by the methodology employed here. Indeed, in both such cases, the
form of the $b_{ab}^{(RF)}(r)$ is important, and this term cannot
obviously be neglected. We have no clue as to how to address situation
where random fluctuations are important part of the physics of the
system, and this remains an important subject for investigations in
the IET methodology.

The separation of the many body correlations into contributions from
frozen fluctuations and those random fluctuations should generally
allow to reproduce the correlations complex soft-matter systems. For
example, micro-emulsions, with the formation of micelles is a first
target for extending IET techniques into soft matter liquids.

\section{Conclusion}

The results shown herein demonstrate that one could extend the HNC
IET in a reliable way deep into the strongly micro-heterogeneous mixtures.
This is accomplished by taking into account specific correlations
which produce this micro-heterogeneity directly into the pair-interaction.
In this work, we have explicitly designed simplified systems, which
incorporate directly into the pair interactions the features which
produce the micro-heterogeneity, which is why we could solve the IET
for such systems. The search for such simplified models indicate that
taking into account the fluctuations which produce the heterogeneity
gives important insights into the nature of the underlying many body
correlations. The division into frozen and fluctuating contribution,
allows to both test the accuracy of the HNC closure, and also appreciate
the role of fluctuations in this closure. The extension of this methodology
to realistic mixtures is the next step. Although the approach remains
empirical, we expect that some general schemes would emerge, in particular
for mixtures involving hydrogen bonding molecules. While the principal
challenge remains to capture the frozen fluctuations which produce
the micro-heterogeneity, the role of the random fluctuations would
assess the reliability of the IET and the HNC approximation. Our expectation
is that, far from phase transition, this methodology should produce
good results since random part of fluctuations are supposed to be
small. In any case, this approach will certainly help to appreciate
the relative importance of both these types of many body correlations.

\section*{Acknowledgments}

A. B. thanks Laboratoire de Physique Théorique de la Matière Condensée
for his first year Master internship.

\newpage{}

\section*{Figure captions}
\begin{lyxlist}{00.00.0000}
\item [{Fig.1}] Pair correlation functions (a) and corresponding structure
factors (b) for the model NA-SRLA, for parameters $\epsilon_{0}=1$,
$\lambda_{0}=1$, $\epsilon_{12}=2.8$, $\lambda=1$, $\epsilon_{1}=40$,
$\alpha_{1}=0.75$, $\kappa_{1}=0.2582$, and for a temperature of
$T=1$ and packing fraction $\eta=0.4$. HNC results in full lines
(blue for 11 component pair and dark green for 12) and simulation
in dashed (cyan for 11 component pair and green for 12). The inset
in (a) represents the effective pair interactions with full lines
for 11 (blue) and 12 (red) and the details for 12 are shown in dotted
lines( LJ part in magenta and Gaussian part in cyan). The inset in
(b) represents the BT structure factors (HNC results for $S_{NN}(k)$
and $S_{CC}(k)$ in full blue and dark green lines, respectively,
simulations results in dashed cyan and green lines, respectively) 
\item [{Fig.2}] Same as Fig.1, but the repulsive interaction for 11 components
and for parameters $\epsilon_{0}=1$, $\lambda_{0}=0$, $\epsilon_{12}=1$,
$\lambda=1$, $\epsilon_{1}=4$, $\alpha_{1}=0.85$, $\kappa_{1}=0.4663$
and same state parameters as in Fig.1. (upper inset in (b) is a zoom
on the pre-peak area). 
\item [{Fig.3}] Pair correlation functions (a) and corresponding structure
factors (b) for the model G-SRLA, for parameters $\epsilon_{0}=\epsilon_{12}=1$,
$\epsilon_{1}=0.5$, $\alpha_{1}=1.5$, $\kappa_{1}=0.41$,$\alpha_{2}=3$,
$\kappa_{2}=0.41$. The lines and color codes are as in Fig.1 
\item [{Fig.4}] Pair correlation functions (a) and corresponding structure
factors (b) for the model G-SRLA, for the same interaction parameters
as in Fig.3, except for $\epsilon_{2}=0.030$. The lines and color
codes are as in Fig.1 
\item [{Fig.5}] Pair correlation functions (a) and corresponding structure
factors (b) for the model G-SRLA, for the same interaction parameters
as in Fig.3, except for $\epsilon_{2}=0.12$. The lines and color
codes are as in Fig.1 
\item [{Fig.6}] Pair correlation functions (a) and corresponding structure
factors (b) for the model G-SRLA, for for the same interaction parameters
as in Fig.3, except for $\epsilon_{2}=0.025$ and $\kappa_{2}=4.47$.
The lines and color codes are as in Fig.1. (upper inset in (b) is
a zoom on the pre-peak area). 
\end{lyxlist}
\newpage{}

.

\includegraphics[scale=0.22]{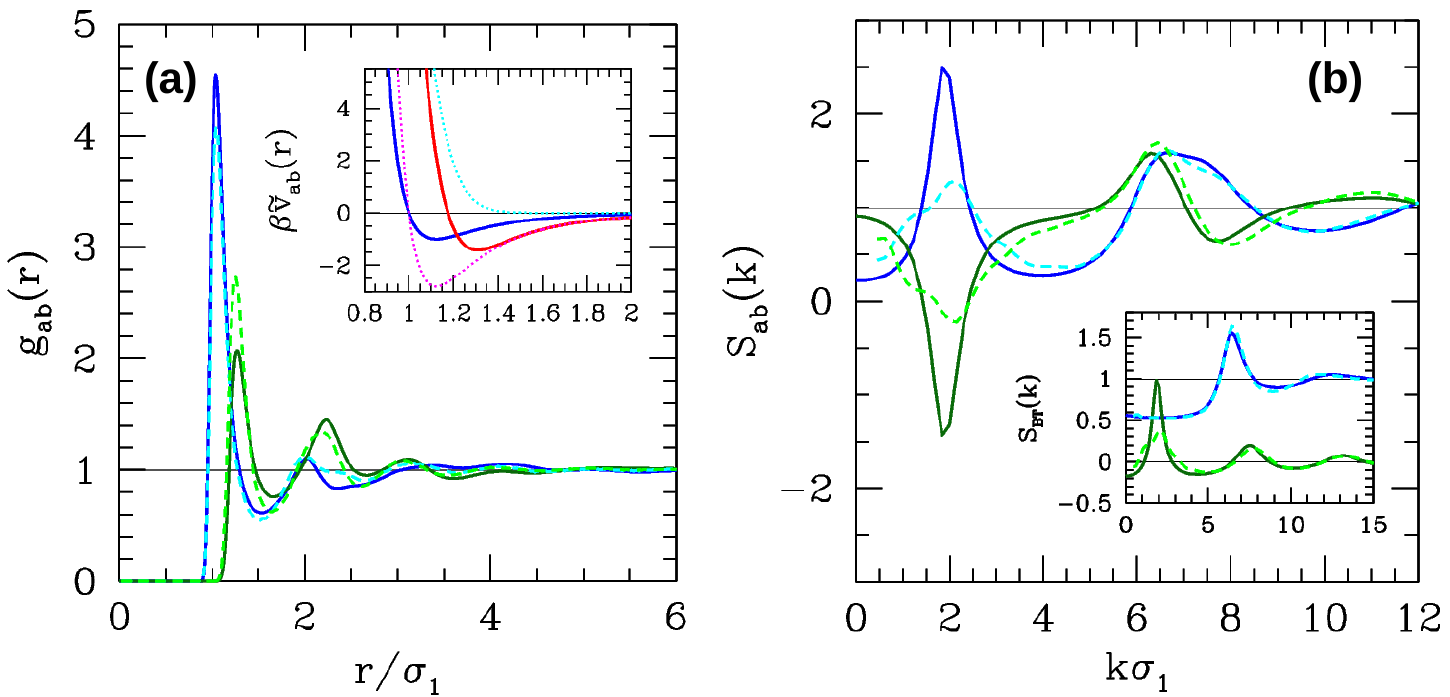}

.

Fig.1 - Pair correlation functions (a) and corresponding structure
factors (b) for the model NA-SRLA, for parameters $\epsilon_{0}=1$,
$\lambda_{0}=1$, $\epsilon_{12}=2.8$, $\lambda=1$, $\epsilon_{1}=40$,
$\alpha_{1}=0.75$, $\kappa_{1}=0.2582$, and for a temperature of
$T=1$ and packing fraction $\eta=0.4$. HNC results in full lines
(blue for 11 component pair and dark green for 12) and simulation
in dashed (cyan for 11 component pair and green for 12). The inset
in (a) represents the effective pair interactions with full lines
for 11 (blue) and 12 (red) and the details for 12 are shown in dotted
lines( LJ part in magenta and Gaussian part in cyan). The inset in
(b) represents the BT structure factors (HNC results for $S_{NN}(k)$
and $S_{CC}(k)$ in full blue and dark green lines, respectively,
simulations results in dashed cyan and green lines, respectively)

.\newpage{}

.

\includegraphics[scale=0.22]{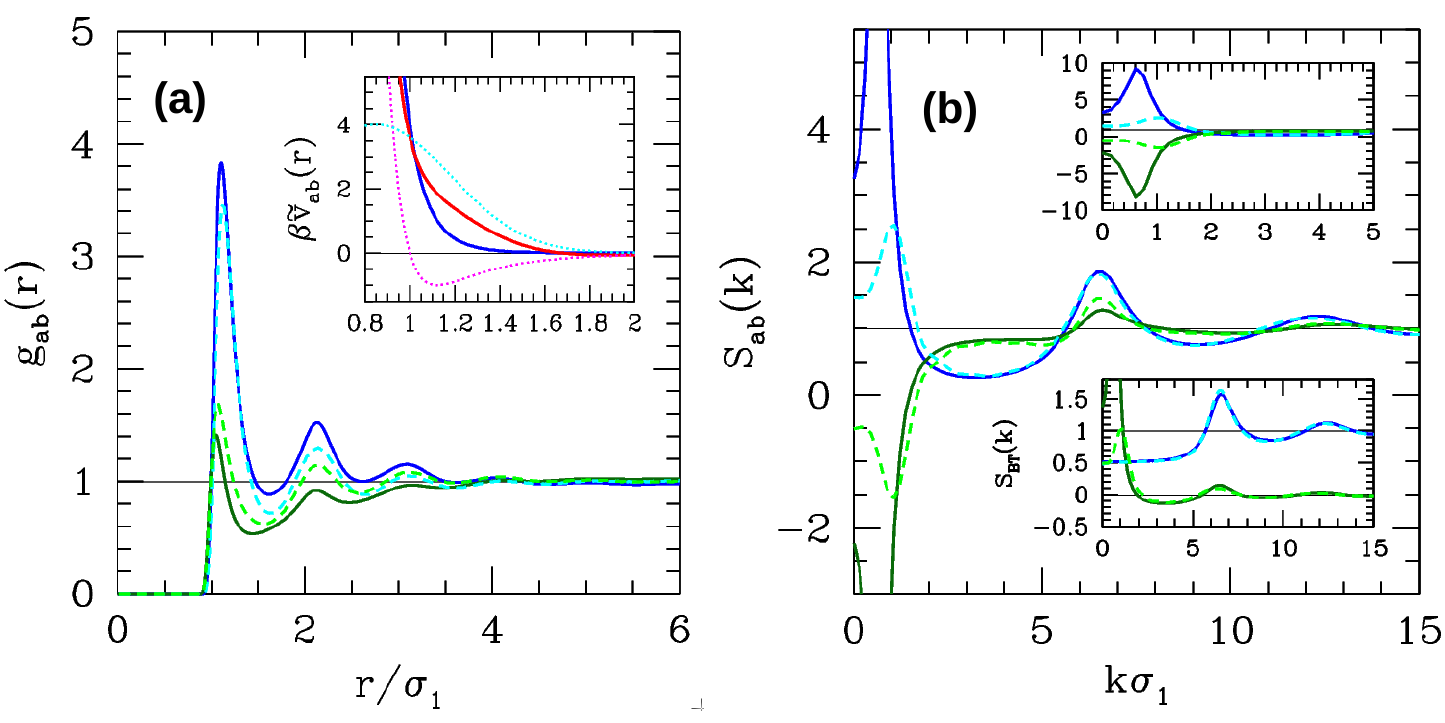}

.

Fig.2 - Same as Fig.1, but the repulsive interaction for 11 components
and for parameters $\epsilon_{0}=1$, $\lambda_{0}=0$, $\epsilon_{12}=1$,
$\lambda=1$, $\epsilon_{1}=4$, $\alpha_{1}=0.85$, $\kappa_{1}=0.4663$
and same state parameters as in Fig.1. (upper inset in (b) is a zoom
on the pre-peak area)

.\newpage{}

.

\includegraphics[scale=0.22]{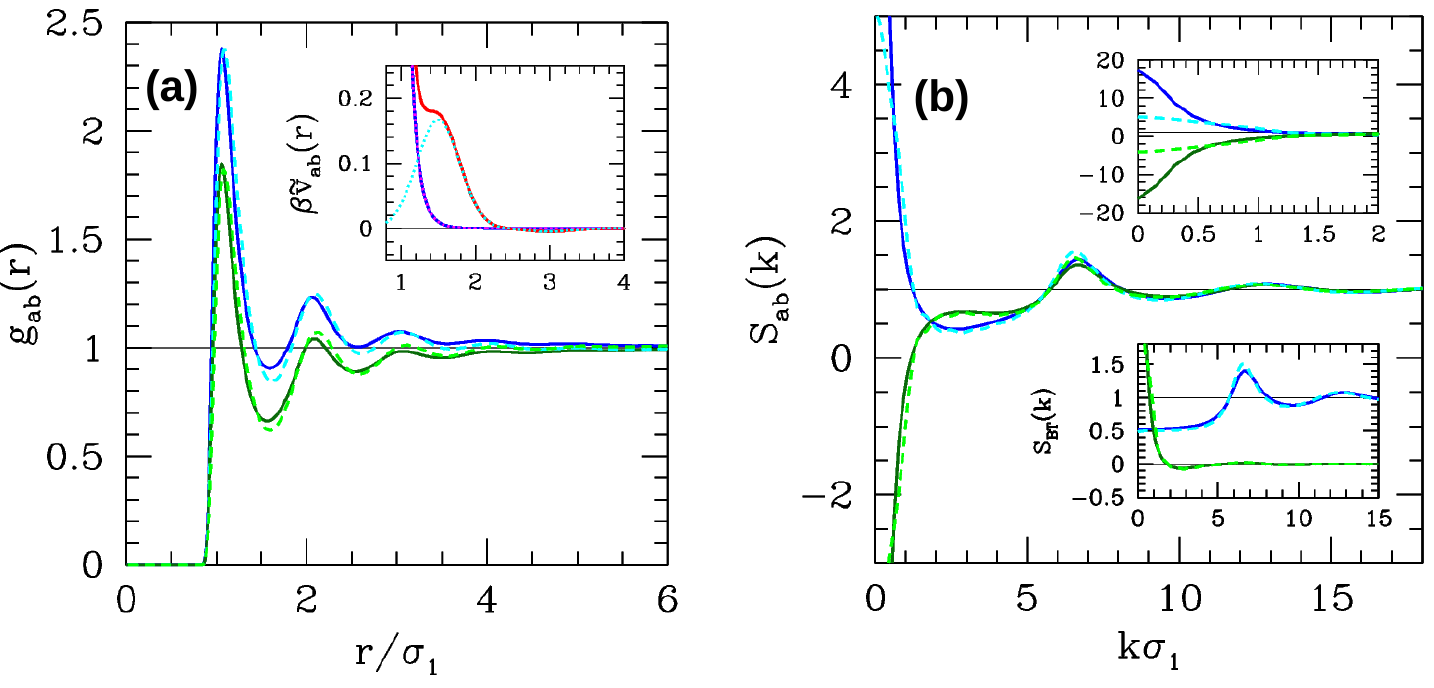}

.

Fig.3 - Pair correlation functions (a) and corresponding structure
factors (b) for the model G-SRLA, for parameters $\epsilon_{0}=\epsilon_{12}=1$,
$\epsilon_{1}=0.5$, $\alpha_{1}=1.5$, $\kappa_{1}=0.41$,$\alpha_{2}=3$,
$\kappa_{2}=0.41$. The lines and color codes are as in Fig.1

.

\newpage{}

.

\includegraphics[scale=0.22]{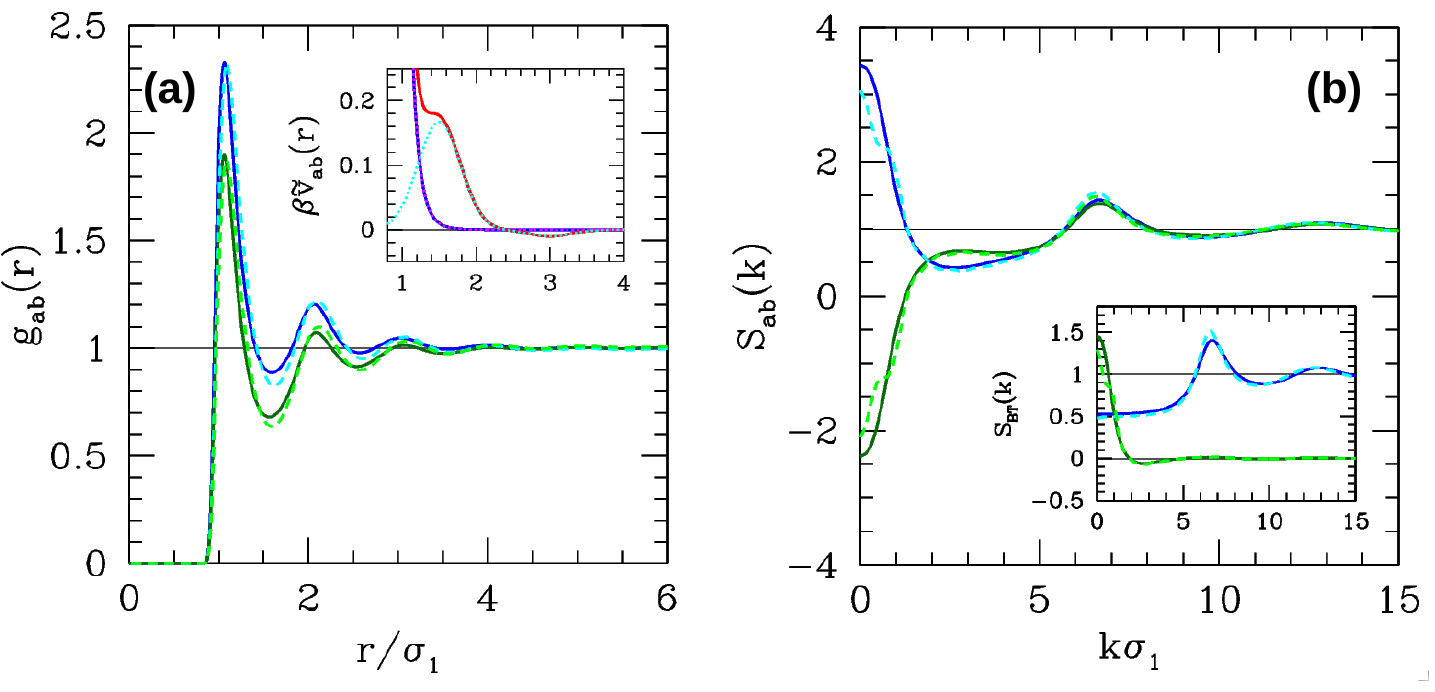}

.

Fig.4 - Pair correlation functions (a) and corresponding structure
factors (b) for the model G-SRLA, for the same interaction parameters
as in Fig.3, except for $\epsilon_{2}=0.030$. The lines and color
codes are as in Fig.1

.

\newpage{}

.

\includegraphics[scale=0.22]{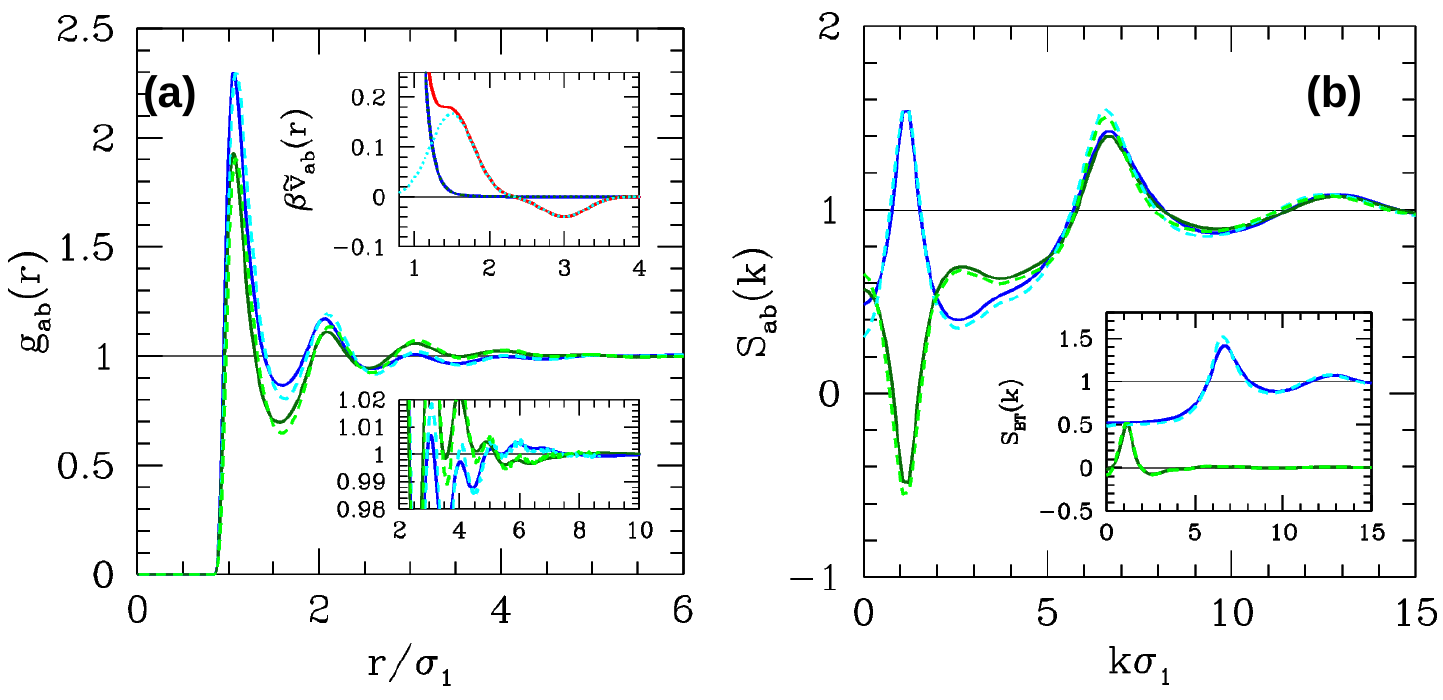}

.

Fig.5 - Pair correlation functions (a) and corresponding structure
factors (b) for the model G-SRLA, for the same interaction parameters
as in Fig.3, except for $\epsilon_{2}=0.12$. The lines and color
codes are as in Fig.1

.

\newpage{}

.

\includegraphics[scale=0.22]{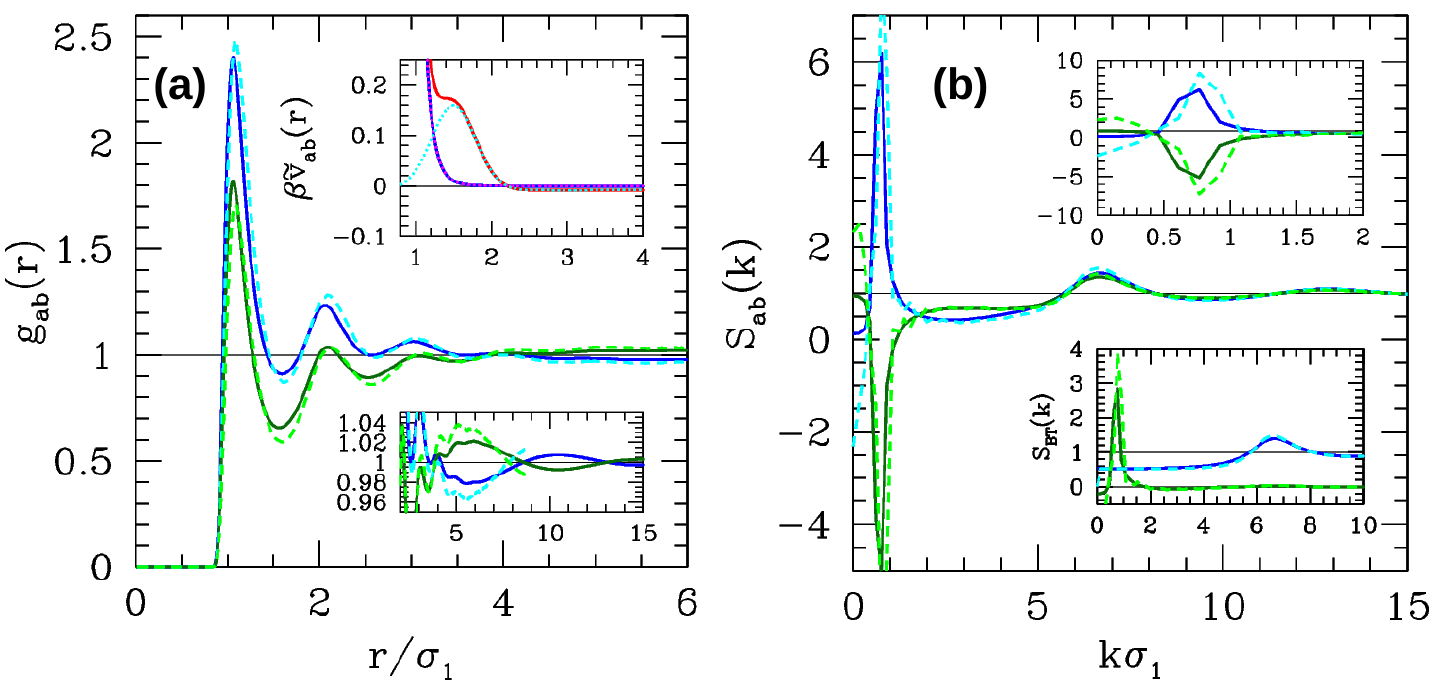}

.

Fig.6 - Pair correlation functions (a) and corresponding structure
factors (b) for the model G-SRLA, for for the(upper inset in (b) is
a zoom on the pre-peak area) same interaction parameters as in Fig.3,
except for $\epsilon_{2}=0.025$ and $\kappa_{2}=4.47$. The lines
and color codes are as in Fig.1.

. 
\end{document}